\documentclass[12pt]{article} 
\usepackage{amsmath}
\usepackage{amssymb}
\usepackage{amstext}
\usepackage{bm}
\usepackage{graphicx}
\usepackage{epstopdf}
\usepackage{cite}

\allowdisplaybreaks 

  \newcommand{\bl}{\big<}
  \newcommand{\bg}{\big>}

\begin{document}
\baselineskip= 0.225in
\parindent=0.35in


\vspace*{0.2in}
\begin{center} 
{\bf \Large Non-classical transport with\\
angular-dependent path-length distributions.\\
2: Application to pebble bed reactor cores\\}
\vspace {0.2in} 
Richard Vasques \\
RWTH University, Aachen, Germany \\ 
Department of Mathematics\\
Center for Computational Engineering Science\\
vasques@mathcces.rwth-aachen.de \\
\vspace {0.1in} 
and \\
\vspace {0.1in} 
Edward W.\ Larsen \\ 
University of Michigan, Ann Arbor, U.S.A. \\
Department of Nuclear Engineering and Radiological Sciences \\ 
edlarsen@umich.edu \\
\end{center}


\vspace{5pt}
\begin{center}
{\bf ABSTRACT}
\end{center}
\begin{quote}
\begin{small}
\vspace{5pt}

We describe an analysis of neutron transport in the interior of model pebble bed reactor (PBR) cores, considering both crystal and random pebble arrangements. Monte Carlo codes were developed for (i) generating random realizations of the model PBR core, and (ii) performing neutron transport inside the crystal and random heterogeneous cores; numerical results are presented for two different choices of material parameters.
These numerical results are used to investigate the anisotropic behavior of neutrons in each case and to assess the accuracy of estimates for the diffusion coefficients obtained with the diffusion approximations of different models: the atomic mix model, the Behrens correction, the Lieberoth correction, the generalized linear Boltzmann equation (GLBE), and the new GLBE with angular-dependent path-length distributions. This new theory utilizes a non-classical form of the Boltzmann equation in which the locations of the scattering centers in the system are correlated and the distance-to-collision is not exponentially distributed; this leads to an anisotropic diffusion equation. We show that the results predicted using the new GLBE theory are extremely accurate, correctly identifying the anisotropic diffusion in each case {\em and} greatly outperforming the other models for the case of random systems.

\end{small} 
\end{quote}

\setlength{\baselineskip}{0.225in}


\vspace{10pt}
\noindent
{\bf 1. INTRODUCTION}
\setcounter{section}{1}
\setcounter{equation}{0} 
\vspace{10pt}

The pebble bed reactor (PBR), a concept which originated in Germany in the 1950’s, is a graphite-moderated, helium-cooled, very-high-temperature (generation IV) reactor. Several countries have addressed different possible PBR designs, amongst which we mention the HTR-PM \cite{zhang_04} in China (following the successful test reactor HTR-10 \cite{wu_02}), the PBMR \cite{koster_03} in South Africa, and the MPBR \cite{kadak_07} in the United States.

The fundamental PBR design is based on the use of spherical, same-sized fuel elements called pebbles. Each fuel pebble is made of pyrolytic graphite (the moderator), containing $\approx$10,000 microscopic fuel Tristructural-Isotropic (TRISO) particles. 
To achieve the desired reactivity, thousands of pebbles are piled on top of one another in a``random" manner (influenced by gravity) inside the cylindrical reactor core. They are dropped on top of this piling from charging tubes located at the top of the core, and move downward through the system in a dense granular flow.
Due to this dynamic structuring, the exact locations of the pebbles inside the core at any given time are unknown. 

Typically, the neutronic modeling of the geometrically random core is done by: (i) developing self-shielded multigroup cross sections for the pebbles, (ii) volume-averaging these cross sections over the entire core, including the helium-filled region between the pebbles (the {\em atomic mix} approximation), and (iii) introducing the spatially-homogenized cross sections into a diffusion code. This procedure leads to two concerns, as explicited next. 

First, in the classic atomic mix model, the cross sections for a random heterogeneous medium are approximated by volume-averaging over the constituent materials, weighted by their respective volume fractions. This approximation is known \cite{larsen_05,vasques_05} to be accurate only when the chunk sizes of the constituent materials are small compared to a mean free path; however, the pebbles are O(1) mean free paths thick. In fact, it has been observed that neutron
streaming in this type of system is strongly affected by the void spaces; to account for this effect, experimental and mathematical approaches were used to develop corrections for the diffusion coefficients obtained with atomic mix \cite{behrens_49,lieberoth_80}. Hence, the validity of the atomic mix approximation is called into question.

The second concern is related but subtly different: in a PBR core, does gravity affect the distance-to-collision in a direction-dependent (anisotropic) manner? In other words, the force of gravity (let us say it acts in the negative $z$ direction) causes the pebbles to arrange themselves in a certain manner, which is affected by the direction of this force. If one considers a typical arrangement of pebbles in a PBR core, the question is whether the chord length probability distribution function in the $z$ direction is different than in the $(x, y)$-plane.

Currently, PBR cores are modeled using a diffusion approximation with an isotropic diffusion tensor, in which neutrons diffuse equally in all spatial directions. Previous research \cite{mathews_93,vasques_09} has indicated that, for different crystal arrangements of a PBR core, anisotropic diffusion effects can be found. In this work, we address this anisotropic behavior and also investigate its existence in PBR random structures.

The basic idea consists of using an angular-dependent, non-exponential ensemble-averaged probability distribution function to replace the true probability distribution function for the distance-to-collision. This leads to the new generalized linear Boltzmann equation (GLBE) derived in \cite{arxiv_13}, and to its asymptotic diffusion limit: a diffusion equation with anisotropic diffusion coefficients. To investigate the accuracy of this extended theory, we have performed numerical simulations in model PBR cores, and compared the diffusion coefficients obtained  numerically with those estimated by the atomic mix model (and its corrections) and by the new generalized theory. Overall, we find that the new theory yields extremely accurate results, correctly predicting anisotropic diffusion in each case and, more importantly, greatly outperforming the other models for the problems in random systems.

A summary of the remainder of the paper follows. In Section 2 we present the different formulations that model neutron transport in the PBR systems considered in this paper, and discuss how to obtain the diffusion coefficients. In Section 3 we show how the crystal and random realizations of the PBR model core were constructed. In Section 4 we discuss the Monte Carlo algorithm used to model neutron transport, and present the numerical results. In Section 5 we compare the theoretically estimated diffusion coefficients with the ones obtained numerically; and in Section 6 we conclude with a discussion.

\vspace{10pt}
\noindent
{\bf 2. FORMULATIONS}
\setcounter{section}{2}
\setcounter{equation}{0} 
\vspace{10pt}

The goal of this paper is to assess the accuracy of the new generalized theory in predicting the diffusion of neutrons in the \textit{interior} of a PBR system; that is, away from boundary effects and strong packing fluctuations. In this section, we shortly describe the different formulations that model neutron transport and diffusion inside such systems, and present the expressions to estimate the diffusion coefficients.

Let the position vector $\bm x$ be described by the cartesian coordinates $x,y,z$, and let us write as $\theta$ the polar angle measured with respect to the (vertical) $z$-axis and as $\varphi$ the corresponding azimuthal angle. Introducing $\mu=cos(\theta)$, we can write the vector
\begin{equation}\label{2.1}
\bm\Omega=\big(\sqrt{1-\mu^2}\cos(\varphi),\sqrt{1-\mu^2}\sin(\varphi),\mu\big)=\textrm{direction of flight}.
\end{equation}

Now, consider a binary system composed of solid fuel (spherical) pebbles (material 1)
immersed in a void background (material 2). In this case, for $i\in\{1,2\}$,
we write the parameters:
\begin{subequations}\label{2.2}
\begin{align}
\Sigma_{t,i} &= \textrm{total cross section of material $i$},\\
c_i &= \textrm{scattering ratio of material $i$}.
\end{align}
\end{subequations}
Finally, defining
\begin{align}\label{2.3}
s = &\textrm{ the path-length traveled by the neutron since}\\
&\textrm{ its previous interaction (birth or scattering),}\nonumber
\end{align}
we make the following assumptions:
\begin{itemize}
\item[$\mathbf{A_1}$] The physical system is both infinite and statistically homogeneous.
\item[$\mathbf{A_2}$] The system has azimuthal symmetry. (The probability distribution function
for distance-to-collision depends only upon the polar angle.)
\item[$\mathbf{A_3}$] Neutron transport is monoenergetic. 
\item[$\mathbf{A_4}$] Neutron transport is driven by a specified point source located at the origin and isotropically emitting $Q$ neutrons per second.
\item[$\mathbf{A_5}$] The neutron flux $\rightarrow 0$ as $|\bm x|\rightarrow \infty$.
\item[$\mathbf{A_6}$] The ensemble-averaged total cross section $\Sigma_t(\bm\Omega,s)$, defined as
\begin{equation}
\Sigma_t(\bm\Omega,s)ds =  \begin{array}{l}
\text{ the probability (ensemble-averaged over all physical realiza-}\\
\text{ tions) that a neutron, scattered or born at any point $\bm x$ and}\\
\text{ traveling in the direction $\bm\Omega$, will experience a collision}\\
\text{ between $\bm x + s\bm\Omega$ and $\bm x + (s+ds)\bm\Omega$,}\\
\end{array}\nonumber
 \end{equation}
is known. 
\item[$\mathbf{A_7}$]  Scattering is isotropic.
\end{itemize}
Assumption $\mathbf{A_2}$ follows directly from the fact that, for any given PBR system (crystal or random), a realization containing $N$ pebbles with coordinates $(x_n, y_n, z_n)$ and a realization containing $N$ pebbles with coordinates $(y_n, x_n, z_n)$, where $1\le n\le N$, have the same probability of occurring.
Therefore, when investigating the behavior of neutrons born in a fuel pebble in the interior of the system, we can assume without loss of generality that the mean-squared displacements in the $x$ and $y$ directions are the same.

\vspace{5pt}
\noindent
{\bf 2.1 The Atomic Mix Model}
\vspace{5pt}

Given a single realization of the system, we can write the packing fraction of material 1 as
\begin{equation}\label{2.4}
\Gamma=\frac{\textrm{total volume occupied by material 1 in the system}}{\textrm{total volume of system}}.
\end{equation}
Then, defining the operator $\big<\cdot\big>$ as the ensemble average over all possible realizations of the system,
the atomic mix parameters are given by
\begin{subequations}\label{2.5}
\begin{align}
\big<\Sigma_t\big> &= \textrm{volume-averaged total cross section} = \big<\Gamma\big>\Sigma_{t,1}+\big(1-\big<\Gamma\big>\big)\Sigma_{t,2};\\
\big<c\Sigma_t\big> &= \textrm{volume-averaged scattering cross section} = \big<\Gamma\big>c_1\Sigma_{t,1}+\big(1-\big<\Gamma\big>\big)c_2\Sigma_{t,2}.
\end{align}
\end{subequations}
The steady-state atomic mix equation \cite{larsen_05} for this system is 
\begin{equation}\label{2.6}
\bm\Omega\cdot\bm\nabla\big<\psi(\bm x,\bm\Omega)\big>+\big<\Sigma_t\big>\big<\psi(\bm x,\bm\Omega)\big>=
\frac{\big<c\Sigma_t\big>}{4\pi}\big<\Phi(\bm x)\big> + \frac{Q}{4\pi}\delta(x)\delta(y)\delta(z),
\end{equation}
where $\big<\Phi(\bm x)\big> = \int_{4\pi}\big<\psi(\bm x,\bm\Omega)\big>d\Omega$. Defining $\big<\Sigma_a\big>=\big<\Sigma_t\big>-\big<c\Sigma_t\big>$,
the asymptotic diffusion approximation for Eq.\ \eqref{2.6} is
\begin{subequations}\label{2.7}
\begin{align}
-\textrm{D}^{am}\,\nabla^2\big<\Phi(\bm x)\big> + \big<&\Sigma_a\big>\big<\Phi(\bm x)\big>=Q\delta(x)\delta(y)\delta(z),\label{2.7a}\\
&\textrm{D}^{am}=\frac{1}{3\big<\Sigma_t\big>},\label{2.7b}
\end{align} 
\end{subequations}
where D$^{am}$ is the atomic mix diffusion coefficient. Notice that this model assumes classical transport, in which the probability distribution function of the
distance traveled between collisions is an exponential. In this case, the mean and mean-squared free paths are given respectively by $\big<s\big> = 1/\big<\Sigma_t\big>$ and
$\big<s^2\big> = 2\big<s\big>^2$.

\vspace{5pt}
\noindent
{\bf 2.2 Corrections to the Atomic Mix Diffusion Coefficient}
\vspace{5pt}

In 1949, Behrens investigated the increase in the migration length of neutrons in a reactor caused by the presence of ``holes" in the reactor \cite{behrens_49}. In that work, ``holes" are primarily understood to be the coolant spaces, due to the low density of the substances used as coolants. He also noticed that a small anisotropic effect occurred depending on the shape of the holes. For the case of pebble beds in which $r\Sigma_t << 1$, he proposed that the isotropic diffusion coefficient D$^B$ be given by
\begin{align}\label{2.8}
\text{D}^B = \left(1 + \frac{2}{3}\frac{\phi^2}{(1+\phi)^2}r\Sigma_t \text{q}_B\right)\text{D}^{am}\,,
\end{align}
where $\Sigma_t$ is the total cross section of the solid material (pebbles), $r$ is the radius of a pebble, $\phi$ is the hole/material volume ratio of the system, D$^{am}$ is given by \eqref{2.7b}, and q$_B$ is the quotient of the mean-squared path-length through the hole divided by the square of the mean path-length through the hole, estimated by 
\begin{align}\label{2.9}
\text{q}_B = 1+ \frac{1}{8\phi^2}\,.
\end{align}

 Later work \cite{neef_74,scherer_74}
emphasized the general validity of these equations for pebble bed problems. In 1980, Lieberoth \& Stojadinovi\'c \cite{lieberoth_80} revisited this theory. They developed a mock-up model of a pebble
bed using steel balls and measured the coordinates of 3,024 sphere centers so that Monte Carlo games for neutron diffusion could be established. Using these results (as well as Monte Carlo calculations for crystal structures), they proposed an improved expression for q$_B$:
\begin{align}\label{2.10}
\text{q}_L = 1.956+\frac{1}{260\phi^2}\,,
\end{align}
which also more closely relates to the theoretical value q = 2 obtained for randomly overlapping spheres \cite{strieder_68}.
Moreover, under the assumption that no correlation exists between the passage lengths in the holes and in the balls, they developed the following formula for the diffusion lengths:
\begin{align}\label{2.11}
\text{D}^L = \bigg\{1 + &\frac{\phi^2}{(1+\phi)^2}\bigg[\frac{2}{3}r\Sigma_t \text{q}_L 
\\&\quad\quad +
\frac{4}{3}r\Sigma_t\left(\frac{2r^2\Sigma_t^2}{2r^2\Sigma_t^2-1+(1+2r\Sigma_t)e^{-2r\Sigma_t}}-1\right)-1\bigg]\bigg\}\text{D}^{am}\,.\nonumber
\end{align}
This correction is used when more accurate estimates of neutron streaming in pebble bed type reactors are required \cite{bernnat_03,williams_01}.

\vspace{5pt}
\noindent
{\bf 2.3 The New GLBE}
\vspace{5pt}

Let us assume that the probability $p$ that a neutron
will experience a collision while traveling an incremental distance $ds$ in a direction $\bm\Omega$
depends on $\bm\Omega$ and $s$; that is, $p(\bm\Omega,s) = \Sigma_t(\bm\Omega,s)ds$.
Following \cite{arxiv_13}, the new GLBE for this system is given by
\begin{align}
\frac{\partial\psi}{\partial s}&(\bm x,\bm\Omega,s)+\bm\Omega\cdot\bm\nabla\psi(\bm x,\bm\Omega,s)+\Sigma_t(\bm\Omega,s)\psi(\bm x,\bm\Omega,s)\label{2.12}\\
&=\frac{\delta(s)}{4\pi}c\int_{4\pi}\int_0^{\infty}\Sigma_t(\bm\Omega',s')\psi(\bm x,\bm\Omega',s')ds'd\Omega' + \delta(s)\frac{Q}{4\pi}
\delta(x)\delta(y)\delta(z).\nonumber
\end{align}
The asymptotic diffusion approximation for Eq.\ \eqref{2.12} in the case of azimuthal symmetry is written as
\begin{subequations}\label{2.13}
\begin{align}
-\textrm{D}_x^{gt}\frac{\partial^2}{\partial x^2}\Phi(\bm x)-\textrm{D}_y^{gt}\frac{\partial^2}{\partial y^2}\Phi(\bm x)-&\textrm{D}_z^{gt}\frac{\partial^2}{\partial z^2}\Phi(\bm x)+\frac{1-c}{\big<s\big>}\Phi(\bm x) = Q\delta(x)\delta(y)\delta(z),\label{2.13a}
\\ \textrm{D}_x^{gt} = \textrm{D}_y^{gt} &= \frac{1}{8\big<s\big>}\int_{-1}^1[1-\mu^2]s^2_{\bm\Omega}(\mu)d\mu\,\label{2.13b}
\\ \textrm{D}_z^{gt} &= \frac{1}{4\big<s\big>}\int_{-1}^1\mu^2s^2_{\bm\Omega}(\mu)d\mu.\label{2.13c}
\end{align}
\end{subequations}
where D$^{gt}_u$ is the diffusion coefficient in the direction $u$ given by the {\em generalized theory}, $s^2_{\bm\Omega}(\mu)$ is the mean-squared free path of a neutron in the direction $\mu$, and $\bl s\bg$ is the mean free path of a neutron.
It is clear from these equations that the diffusion coefficients given by the new GLBE can differ in vertical and horizontal directions, accounting for anisotropic effects that may be present in the problem.

In the case of $s^2_{\bm\Omega}$ being independent of $\mu$, such that $s^2_{\bm\Omega}(\mu)=\bl s^2\bg$,
the diffusion equation reduces to the one in \cite{larsen_11}:
\begin{subequations}\label{2.14}
\begin{align}
-\textrm{D}^{iso}\,\nabla^2\Phi(\bm x) + \frac{1-c}{\big<s\big>}&\Phi(\bm x)=Q\delta(x)\delta(y)\delta(z),\\
\textrm{D}^{iso}&=\frac{1}{3}\frac{\bl s^2\bg}{2\bl s\bg},\label{2.14b}
\end{align} 
\end{subequations}
where D$^{iso}$ represents a non-classical \emph{isotropic} diffusion coefficient.

\vspace{5pt}
\noindent
{\bf 2.4  Exact Expressions for $\Sigma_a$ and Mean-Squared Displacements}
\vspace{5pt}

Consider the diffusion equation below, taking place in an infinite system with a point source at the origin:
\begin{equation}\label{2.15}
-\textrm{D}_x \frac{\partial^2}{\partial x^2}\Phi(\bm x) -\textrm{D}_y \frac{\partial^2}{\partial y^2}\Phi(\bm x) - \textrm{D}_z \frac{\partial^2}{\partial z^2}\Phi(\bm x)
+ \Sigma_a\Phi(\bm x) = Q(\bm x)\delta(x)\delta(y)\delta(z)\, .
\end{equation}
Bearing in mind that $\Phi(\bm x) \to 0$ and $\nabla\Phi(\bm x)\to 0$ as $|\bm x|\to \infty$, we can manipulate this equation to obtain exact formulas for $\Sigma_a$ and for the mean-squared displacement of neutrons.

Operating on Eq.\ (\ref{2.15}) by
\begin{align}\label{2.16}
\int_{-\infty}^{\infty}\int_{-\infty}^{\infty}\int_{-\infty}^{\infty}(.)dxdydz\,,
\end{align}
we obtain the exact expression
\begin{align}\label{2.17}
\Sigma_a\int_{-\infty}^{\infty}\int_{-\infty}^{\infty}\int_{-\infty}^{\infty}\Phi(\bm x)dxdydz = Q(0)\,.
\end{align}
Bearing in mind that $\Phi(\bm x) dxdydz$ represents the rate at which path-length is generated by neutrons in $dxdydz$ about $\bm x$, we see that
\begin{align}
\int_{-\infty}^{\infty}\int_{-\infty}^{\infty}\int_{-\infty}^{\infty}\Phi(\bm x)dxdydz = &
\begin{array}{l}
  \text{rate at which path-length is} \label{2.18}
\\\text{generated by neutrons in the system}
\end{array}
\\ = &\begin{array}{l}
  \text{[number of neutrons in the system]}
\\\text{$\times$[mean path-length generated by one neutron]}
\end{array} \nonumber
\\ =& \begin{array}{l}
\text{[number of neutrons in the system]}
\\\text{$\times$[mean number of collisions of a neutron]}
\\\text{$\times$[mean free path of a neutron]}
\end{array}\nonumber
\\ =& [Q(0)]\left[\frac{1}{1-c}\right][\langle s\rangle] \,.\nonumber
\end{align}
Introducing this result into Eq.\ \eqref{2.17} we obtain the \textit{exact} expression
\begin{align}\label{2.19}
\Sigma_a = \frac{1-c}{\langle s \rangle}\,.
\end{align}
Notice that for the classic case (in which $\Sigma_t = 1/\langle s \rangle$) this expression reduces to the classic expression $\Sigma_a = (1-c)\Sigma_t$.

Next, multiplying Eq.\ (\ref{2.15}) by $x^2$ and operating on it
by Eq.\ (\ref{2.16}), we obtain:
\begin{subequations}\label{2.20}
\begin{align}\label{2.20a}
-\textrm{D}_x\int_{-\infty}^{\infty}\int_{-\infty}^{\infty}\int_{-\infty}^{\infty}x^2\frac{\partial^2}{\partial x^2}&\Phi(\bm x)dxdydz
\\& +\Sigma_a\int_{-\infty}^{\infty}\int_{-\infty}^{\infty}\int_{-\infty}^{\infty}x^2\Phi(\bm x)dxdydz =0\,.\nonumber
\end{align}
 Integrating the first term on this equation by parts we get
\begin{align}\label{20b}
\int_{-\infty}^{\infty}\int_{-\infty}^{\infty}\int_{-\infty}^{\infty}x^2\frac{\partial^2}{\partial x^2}\Phi(\bm x)dxdydz
= 2\int_{-\infty}^{\infty}\int_{-\infty}^{\infty}\int_{-\infty}^{\infty}\Phi(\bm x)dxdydz\,.
\end{align}
\end{subequations}
 We can rewrite Eq.\ (\ref{2.20a}) as
\begin{align}\label{2.21}
\langle x^2\rangle=\frac{\int_{-\infty}^{\infty}\int_{-\infty}^{\infty}\int_{-\infty}^{\infty}x^2\Phi(\bm x)dxdydz}
{\int_{-\infty}^{\infty}\int_{-\infty}^{\infty}\int_{-\infty}^{\infty}\Phi(\bm x)dxdydz} =
 2\frac{\textrm{D}_x}{\Sigma_a}\,,
\end{align}
where $\langle x^2\rangle$ represents the mean-squared displacement of a neutron \textit{in the $x$ direction}. This argument yields similar results for the $y$ and $z$directions,
giving us an exact expression for the diffusion coefficient in direction $u$ in terms of the mean-squared displacement:
\begin{equation}\label{2.22}
\text{D}_u=\frac{\bl u^2 \bg}{2}\frac{(1-c)}{\bl s\bg}.
\end{equation}

\vspace{10pt}
\noindent
{\bf 3. CONSTRUCTING PBR MODELS}
\setcounter{section}{3}
\setcounter{equation}{0} 
\vspace{10pt}

In this section we discuss the construction of crystal and random systems that model a realization of a PBR core. The cylindrical geometry of the actual core is not addressed, since we are interested in analyzing neutron diffusion in the interior of the system (away from the boundaries).

\vspace{5pt}
\noindent
{\bf 3.1 Crystal Structures}
\vspace{5pt}

\begin{figure}[!ht]
\centering
\includegraphics[scale=0.07,angle=0]{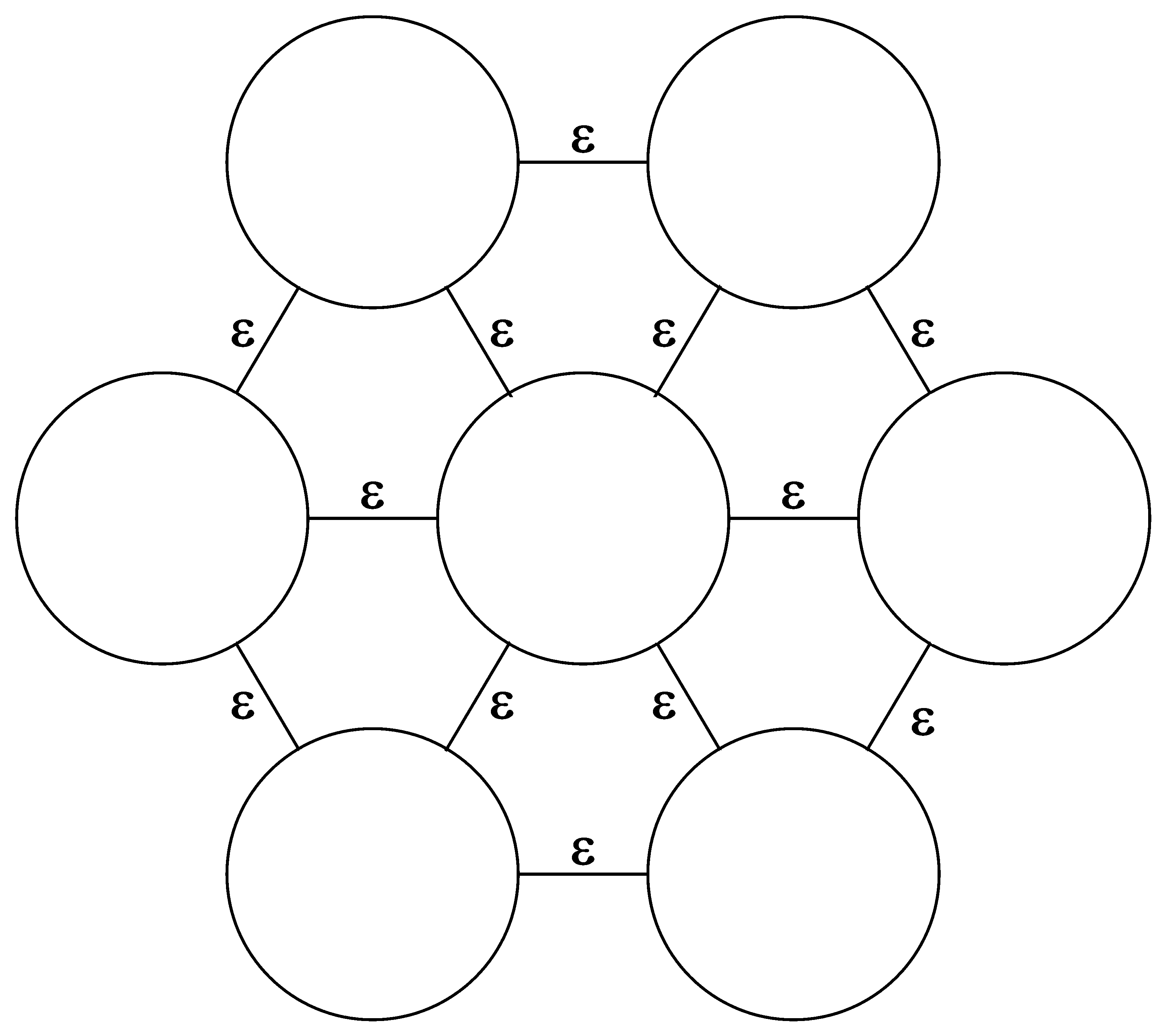}
\caption{Arrangement of pebbles in a horizontal layer with a given distance $\varepsilon$}\label{fig1}
\end{figure}
There are several possible ways of piling identical hard spheres in a crystal-like structure. In particular, it has been shown \cite{hales_05,hales_06,hales_11} that the maximum packing fraction of identical spheres in a container is given by the face-centered cubic arrangement (FCC), with packing fraction $\Gamma\approx 0.74048$. For this reason, we have based the crystal structures in this work in the FCC arrangement.

\begin{figure}[!bht]
\centering
\includegraphics[scale=0.35,angle=0]{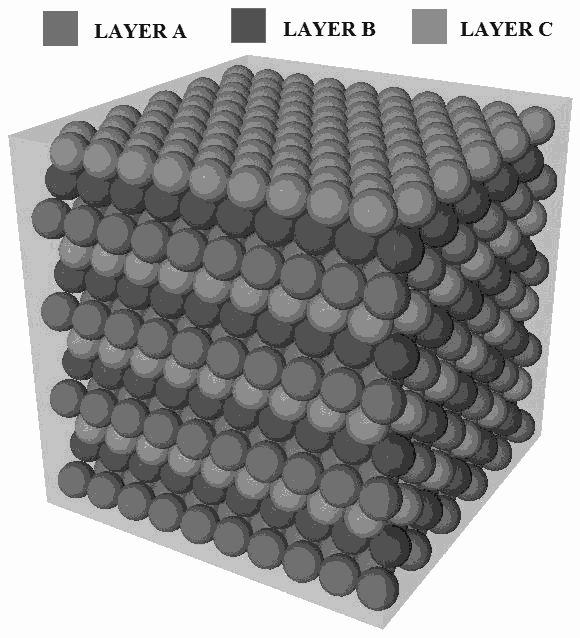}
\caption{FCC crystal structure ($\varepsilon = 0$) in a box with side $L = 10$d}\label{fig2}
\end{figure}

Let d$= 2r$ be the diameter of a fuel pebble, and $\varepsilon$ be the fixed distance between pebbles in the same layer, as shown in Figure \ref{fig1}.
We place the first layer (A) of pebbles at the bottom of the system, and lock them in place. We then proceed to fill the system in a face-centered fashion; that is, positioning the second (B) and third (C) layers and sequentially repeating this structure. A finite example of this type of piling is shown in Figure \ref{fig2}. The height $h_i$ of the $i^{th}$ layer can be defined directly from the previous layers by
\begin{align}\label{3.1}
h_i = h_{i-1}+\sqrt{\text{d}^2-\frac{1}{3}(\text{d} + \varepsilon)^2} = h_{1}+(i-1)\sqrt{\text{d}^2-\frac{1}{3}(\text{d} + \varepsilon)^2}\,,\,\, \forall \,\,i\geq 1\,,
\end{align} 
with $\varepsilon = \text{d}/4 = r/2$.

For $\varepsilon = 0$, this packing method yields the
classic FCC structure, with coordination number (number of spheres contacted by a given sphere) 12.
For the cases with $\varepsilon > 0$, however, the 
\begin{figure}[!ht]
\centering
\includegraphics[scale=0.35,angle=0]{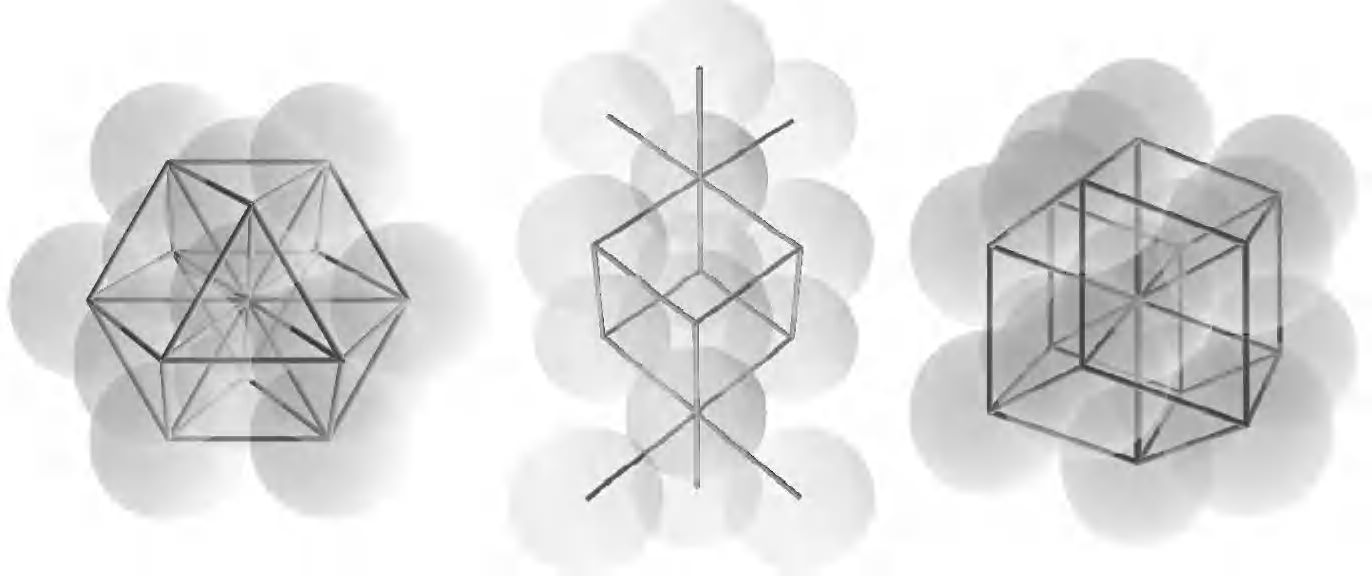} 
\caption{Crystal packings with $\varepsilon = 0$ (left); $0<\varepsilon<\varepsilon_{max}$ (center); $\varepsilon = \varepsilon_{max}$ (right)}\label{fig3}
\end{figure}
``cubic" feature of these face-centered systems is lost; the structure resembles that of a face-centered orthorhombic.
 Moreover, since we do not allow pebbles to overlap each other, $\varepsilon$ must not exceed a maximum value if the crystal-like 
structure of the packing is to be maintained:
\begin{align}\label{3.2}
\varepsilon_{max} = \text{d}\left(\frac{2}{3}\sqrt{6}-1\right)\,.
\end{align}
The packing structures generated with $0<\varepsilon<\varepsilon_{max}$ have coordination number 6.
The packing generated with $\varepsilon = \varepsilon_{max}$ is \textit{not} a rotation of the one generated with $\varepsilon = 0$; it is rather a geometrically different structure, with coordination number 8. A diagram of each case is given in Figure \ref{fig3}.
The graph in Figure \ref{fig4} shows the packing fraction $\Gamma$ as a function of $\varepsilon/\text{d}$ (in increments of 0.025).
\begin{figure}[h]
\centering
\includegraphics[scale=0.45,angle=0]{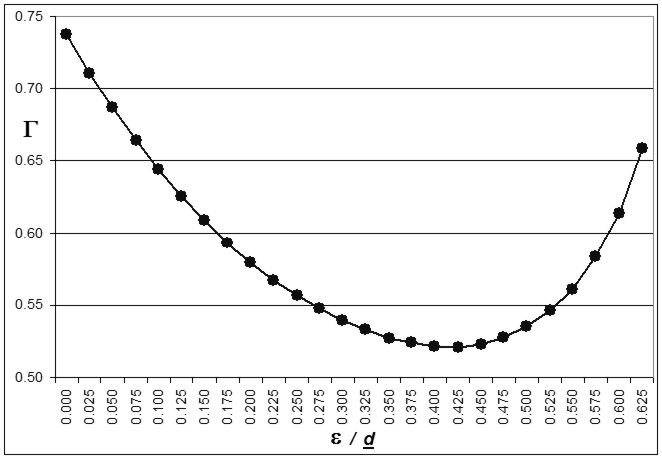}
\caption{Packing fractions in crystal structures with different values of $\varepsilon$}\label{fig4}
\end{figure}

\pagebreak
\vspace{5pt}
\noindent
{\bf 3.2 Random Structures}
\vspace{5pt}

As summarized in \cite{nature_72}, exhaustive experimental work has lead to believe that randomly packed spheres of the same diameter cannot have a packing fraction larger than $\approx 0.64$; in fact, recent analytical work yields the same results \cite{song_08}.
It has also been argued, however, that random packing itself is not well-defined \cite{torquato_00,torquato_02}.

For the problem of PBR cores, the most accepted average packing fraction ranges from 0.60 to 0.62, values that were experimentally validated in \cite{wakil_82}. However, the probability of occurrence of any single packing structure is not quantified, and as pointed out in \cite{ougouag_01}, there exists neither experimental evidence
nor theoretical proof to support the assertion that other packing arrangements within the core are
impossible. In fact, it has been shown \cite{cogliati_06} that changes in the friction coefficients have a significant influence on the structuring of the pebbles; under certain loading circumstances, packing fractions of 0.59 are possible.

To generate the random packings presented in this work, we have used a variation of the ballistic deposition method \cite{jullien_88,meakin_90} introduced in \cite{vasques_09}.
In our ballistic algorithm, each pebble is released at a random point above a cubic box. It then follows a steepest descent trajectory until it reaches a position that is stable under gravity, in which case it has its coordinates stored.
Given a (incomplete) realization of the system, we randomly drop and store the coordinates of 20 different \textit{tentative} pebbles; we then choose the one with the lowest $z$-coordinate to be added to the system, and discard the 19 remaining ones. Once a pebble is added to the system, its position is locked; that is, the pebble is frozen in place.
Rearrangement of pebbles and/or cascading events cannot happen. No velocity or friction coefficients are taken into account; the only restriction is that a pebble can never, at any point of its trajectory, overlap the limits of the box or another pebble.
 Once the tentative pebble with the lowest $z$-coordinate is added to the system, the process is repeated;
 \begin{figure}[!ht]
\centering
\includegraphics[scale = 0.4, angle=0]{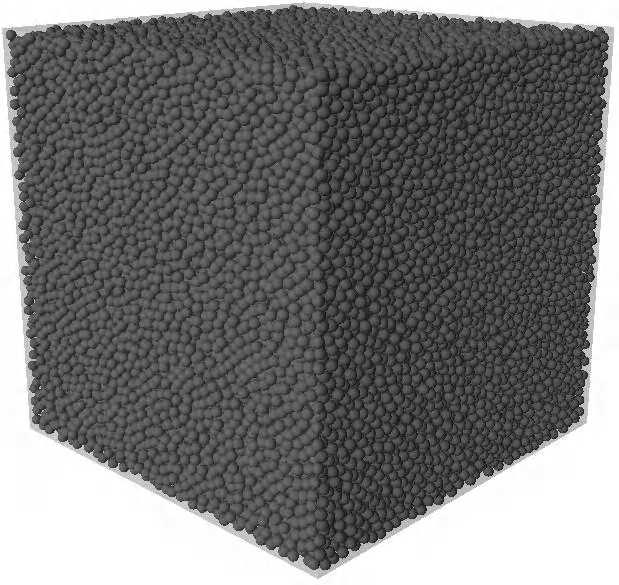}
\caption{Example of a pebble bed random structure in a box with side $L = 50$d}\label{fig5}
\end{figure}
 pebbles continue to be added until the box is full. An example of a random piling obtained with this procedure is shown in Figure \ref{fig5}.   

We have developed 100 different random packings in a box with side $L = 50\text{d}$. Assuming an imaginary box with side $L^*=44\text{d}$ inside the system, such that its walls are a distance $3\text{d}$ away from the walls of the box, we can define the packing fraction $\Gamma^*$ as the ratio between the total volume of pebbles inside this imaginary box (including partial pebbles) and the volume of the imaginary box.
For the 100 simulated random packings in this work, we have found the average packing fraction {\em in the interior of the system} to be $\bl\Gamma^*\bg = 0.5934$, with a standard deviation of 0.0012.

\vspace{10pt}
\noindent
{\bf 4. MONTE CARLO NUMERICAL RESULTS}
\setcounter{section}{4}
\setcounter{equation}{0} 
\vspace{10pt}

Unfortunately, due to the statistical nature of the heterogeneous medium and its effect on the transport of neutrons, there is generally no analytical expression that can be used to obtain the mean and mean-squared free paths in PBR random systems.
Nevertheless, there is a logical and straightforward set of steps that can be followed in order to numerically estimate this quantity.

If we 
consider a particle P that is born (or scatters) at a random point $(x,y,z)$ inside the pebble $S_0$,
the total distance $\hat s$ that this particle will travel
\textit{inside the pebbles} before experiencing a collision
can be sampled from the exponential distribution
\begin{equation}\label{4.1}
 q(\hat s) = \Sigma_{t,1}e^{-\Sigma_{t,1}\hat s},
 \end{equation}
where $\Sigma_{t,1}$ is the total cross section of 
the pebbles.

Let $\ell$ be the linepath starting at point $(x,y,z)$
along which P travels, and let
$\delta_{S_n}$ be the length of $\ell$ inside the pebble $S_n$. If 
$\delta_{S_0} < \hat s$,
 P will leak out of the pebble $S_0$ before experiencing a collision. It will then travel
 \begin{figure}[!ht]
\centering
\includegraphics[scale=0.1,angle=0]{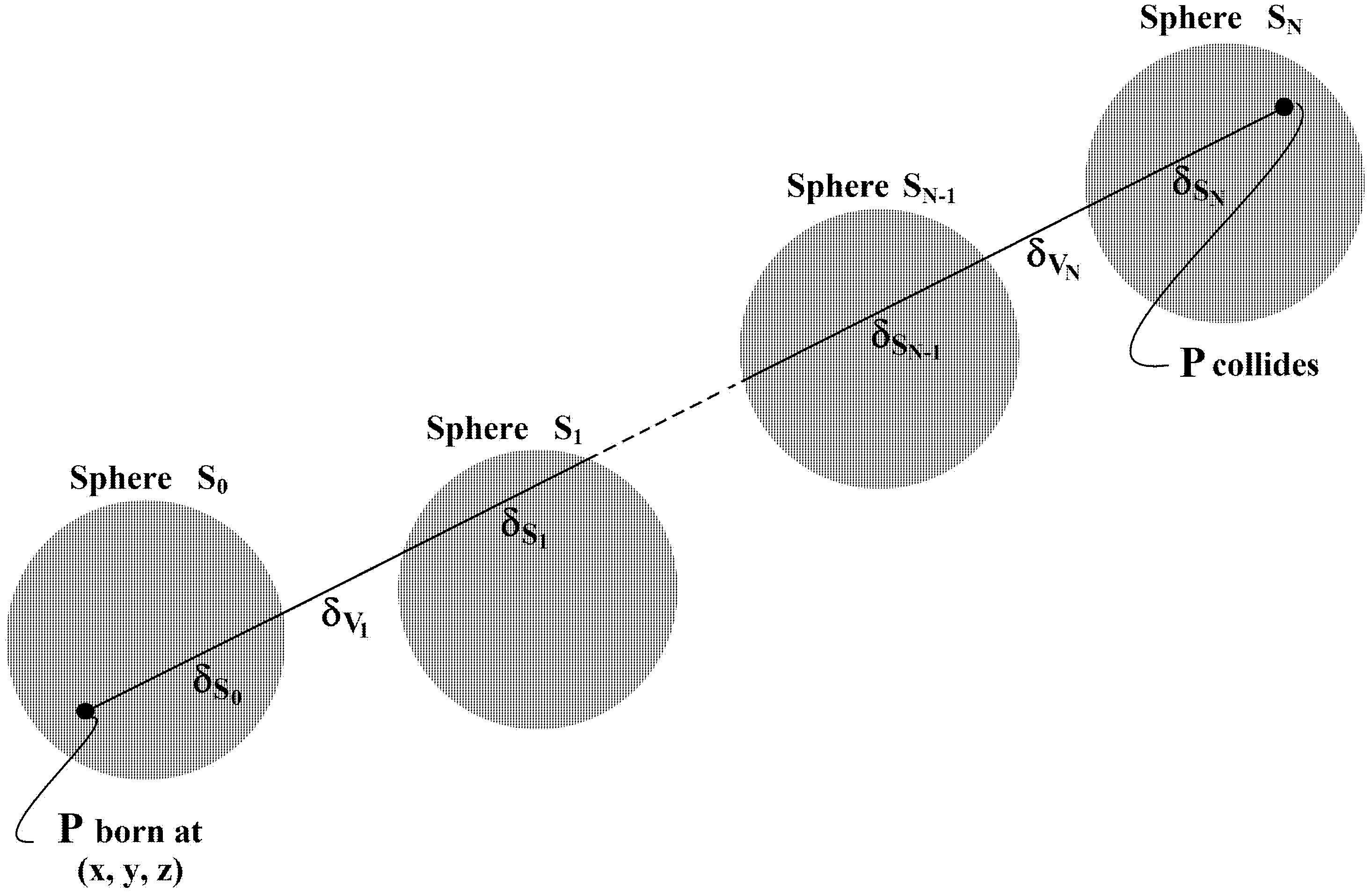}
\caption{Linepath of a particle between collisions}\label{fig6}
\end{figure}
a distance $\delta_{V_1}$ along $\ell$ in the vaccum before entering a new pebble $S_1$. If $\delta_{S_1} < \hat{s}-\delta_{S_0}$,
the particle will leak out of $S_1$ without experiencing a collision and will travel some distance $\delta_{V_2}$ along $\ell$ in the vacuum
before entering another pebble ($S_2$).
 Eventually,
if the particle does not leak out of the system,
there will be a pebble $S_N$ in which 
$\delta_{S_N}\geq \hat{s}-\displaystyle{\sum_{n=0}^{N}\delta_{S_n}}$,
meaning that P will experience a collision within $S_N$ (Figure \ref{fig6}). The distance traveled by this particle
between birth and collision is given by
\begin{align}\label{4.2}
s = \hat{s} + \displaystyle{\sum_{n=1}^{N}\delta_{V_n}}.
\end{align}
In a given realization of this system, we use the following procedure:
\begin{itemize}
\item[(i)] Choose a pebble in the system.\vspace{-8pt}
\item[(ii)] Randomly choose a point inside this pebble.\vspace{-8pt}
\item[(iii)] Using Eqs.\ \eqref{4.1} and (\ref{4.2}), calculate $s$ and $s^2$.\vspace{-8pt}
\item[(iv)] Repeat this process for a large number of particles.
\end{itemize}
We tally the results obtained with this process
in different directions $\mu$, and then divide $\displaystyle{\sum s}$ and $\displaystyle{\sum s^2}$ by the number of linepaths generated to obtain $s_{\bm\Omega}(\mu)$ and $s^2_{\bm\Omega}(\mu)$, the mean and mean-squared distance-to-collision in the fixed direction $\mu$. 
\begin{table}[!ht]
\small
\centering
\caption{Parametes for fuel pebbles with diameter d}\label{tab1}
\begin{tabular}{||c||c|c|c||c|c||}\hline\hline
Problem & d$\Sigma_t  $ & d$\Sigma_s$ & d$\Sigma_a$ & $c=\Sigma_s/\Sigma_t$ & $P(\bm\Omega\cdot\bm\Omega')$\\
\hline
\hline
\textbf{1} &1.0 & 0.99 & 0.01 & 0.99& 1/4$\pi$ \\
\hline
\textbf{2} &2.0 & 0.995 & 0.005 & 0.9975& 1/4$\pi$ \\
\hline\hline
\end{tabular}
\end{table}

For the two sets of parameters considered in this work, we assumed the background material in which the pebbles were piled to be vacuum; the parameters used for the material of the pebbles are given in Table \ref{tab1}.
The neutron histories within the systems are determined by a Monte Carlo transport code as follows:
\begin{itemize}
\item[\textbf{I}] A neutron $n$ is born at a random point inside the fuel pebble.
\item[\textbf{II}] A random direction of flight $\bm\Omega$ is defined.
\item[\textbf{III}] The distance that $n$ travels \textit{inside the pebbles} is sampled from Eq.\ \eqref{4.1}.
\item[\textbf{IV}] $n$ undergoes a collision and all relevant information is stored.
\item[\textbf{V.i}] If $n$ is absorbed, the history of $n$ ends and the algorithm goes back to step \textbf{I}.
\item[\textbf{V.ii}] If $n$ is scattered, the algorithm goes back to step \textbf{II}.
\end{itemize}
In this algorithm, we make use of the azimuthal symmetry of the system to improve our results involving the $(x,y)$-plane. In short, this means that we obtain (without loss of generality) the result $\bl x^2\bg = \bl y^2\bg$.
\begin{table}[!ht]
\small
\centering
\caption{Monte Carlo results in crystal structures with different values of $\varepsilon$}\label{tab2}
\begin{tabular}{||c||c|c|c|c||c|c|c|c||}\hline\hline
&\multicolumn{4}{|c||}{\bf{Problem 1}}&\multicolumn{4}{|c||}{\bf{Problem 2}}\\
\cline{2-9}
&&&&&&&&\\
$\varepsilon/$d & $\bl s\bg/$d & $\bl s^2\bg/$d & $\bl x^2\bg/$d & $\bl z^2\bg/$d &$\bl s\bg/$d & $\bl s^2\bg/$d & $\bl x^2\bg/$d & $\bl z^2\bg/$d \\
&&&&&&&&\\
\hline\hline
0.000&1.3501 &3.7634 &  125.15& 124.96&0.6751 & 0.9728 & 128.90 & 128.84\\
\hline
0.025 & 1.4003 & 4.0685 & 135.27&135.59 & 0.7003 & 1.0570&139.55 & 140.77 \\
\hline
0.050 & 1.4497 & 4.3830 & 145.44&146.05 & 0.7250 & 1.1452 & 150.68 & 153.68\\
\hline
0.075 & 1.4982 & 4.7070 & 156.12&157.30 & 0.7492 & 1.2366 & 162.54 & 166.08 \\
\hline
0.100 & 1.5452 & 5.0358 & 167.00&168.61 & 0.7728 & 1.3308 & 174.15 & 178.79\\
\hline
0.125 & 1.5909 & 5.3687 & 177.51&180.10 & 0.7956 & 1.4268 & 186.53 & 192.58 \\
\hline
0.150 & 1.6349 & 5.7031 & 188.50&192.02 & 0.8176 & 1.5240 & 198.63 & 205.33\\
\hline
0.175 & 1.6769 & 6.0349 & 199.07&202.45 & 0.8387 & 1.6214 & 211.23 & 218.42\\
\hline 
0.200 & 1.7165 & 6.3604 & 209.74&214.16 & 0.8585 & 1.7179 & 223.91 & 231.18\\
\hline
0.225 & 1.7538 & 6.6769 & 220.35&224.06 & 0.8770 & 1.8118 & 235.74 & 243.91 \\
\hline
0.250 & 1.7879 & 6.9768 & 230.37&233.63 & 0.8942 & 1.9026 & 247.41 & 255.24\\
\hline
0.275 & 1.8189 & 7.2588 & 239.58&243.23 & 0.9096 & 1.9876 & 258.88 & 266.86 \\
\hline
0.300 & 1.8459 & 7.5125 & 248.20&251.55 & 0.9233 & 2.0655 & 268.49 & 276.59\\
\hline
0.325 & 1.8690 & 7.7347 & 255.47&258.64 & 0.9348 & 2.1341 & 278.33 & 284.99 \\
\hline
0.350 & 1.8873 & 7.9166 & 261.44&263.55 & 0.9439 & 2.1904 & 286.02 & 290.14 \\
\hline
0.375 & 1.9004 & 8.0490 & 266.50&267.98 & 0.9504 & 2.2315 & 292.04 & 295.23 \\
\hline
0.400 & 1.9074 & 8.1193 & 268.96&269.23 & 0.9540 & 2.2547 & 296.11 & 297.01  \\
\hline
0.425 & 1.9080 & 8.1260 & 269.74&269.10 & 0.9542 & 2.2554 & 296.91 & 295.76\\
\hline
0.450 & 1.9007 & 8.0497 & 268.14&265.50 & 0.9507 & 2.2330 & 294.33 & 290.01 \\
\hline
0.475 & 1.8851 & 7.8904 & 262.72&259.34 & 0.9428 & 2.1825 & 288.24 & 281.70\\
\hline
0.500 & 1.8596 & 7.6390 & 254.66&250.14 & 0.9300 & 2.1034 & 278.96 & 271.24\\
\hline
0.525 & 1.8225 & 7.2841 & 243.20&239.29 & 0.9115 & 1.9942 & 265.10 & 256.42\\
\hline
0.550 & 1.7723 & 6.8263 & 227.98&223.77 & 0.8864 & 1.8552 & 247.07 & 238.01\\
\hline
0.575 & 1.7063 & 6.2597 & 209.48&205.57 & 0.8533 & 1.6854 & 224.80 & 216.64 \\
\hline
0.600 & 1.6209 & 5.5806 & 186.37&183.61  & 0.8106 & 1.4855 & 197.92 & 192.47\\
\hline 
0.625 & 1.5110 & 4.7854 & 159.37&158.03  & 0.7557 & 1.2567 & 166.72 & 165.40\\
\hline
\hline
\end{tabular}
\end{table}

\vspace{5pt}
\noindent
{\bf 4.1 Numerical Results in Crystal Structures}
\vspace{5pt}

For all crystal structures considered in this work, the model core consists of a periodically infinite structure of pebbles of diameter d. The histories of 3$\times 10^6$ neutrons were simulated in each system; Monte Carlo numerical results are given in Table \ref{tab2}. The statistical error
(with 95\% confidence) is less than 0.037$\%$ for all values of $\bl s \bg$, $\bl s^2\bg$, $\bl x^2\bg=\bl y^2\bg$, and $\bl z^2\bg$.
\begin{figure}[!ht]
\centering
\includegraphics[scale=0.5,angle=0]{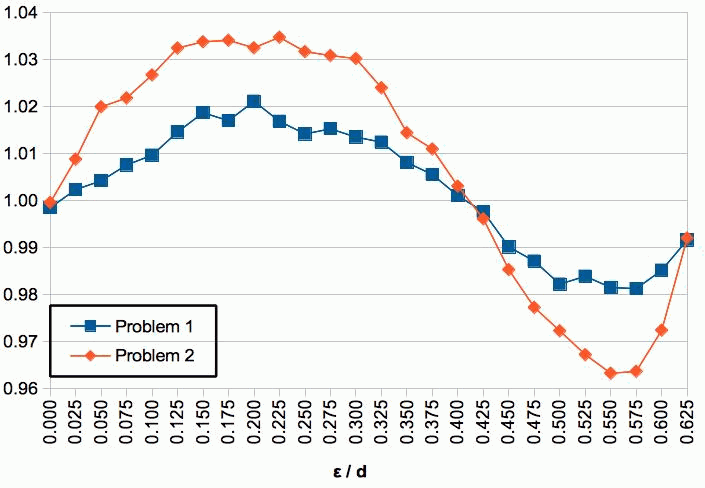} 
\caption{Ratio $\bl z^2\bg/\bl x^2\bg$ in crystal structures as a function of $\varepsilon$}\label{fig7}
\end{figure}

As expected, we detect a clear anisotropic effect in each of these systems, as can be seen by the ratio $\bl z^2\bg/\bl x^2\bg$ depicted in Figure \ref{fig7}. This indicates that the diffusion coefficients must be different for the vertical and horizontal directions.

\vspace{5pt}
\noindent
{\bf 4.2 Numerical Results in Random Structures}
\vspace{5pt}

For all random systems in this work, the packing of pebbles  of diameter d took place in a cubic box with side $L=50$d. We choose the pebble closest to the center of the packing structure to be the one where neutrons are born. Being interested in the behavior of neutrons
in the interior of the system, we want to minimize the effect of the boundaries of the box (walls, top, bottom).
According to the work in
\begin{figure}[!ht]
\centering
\includegraphics[scale=0.35,angle=0]{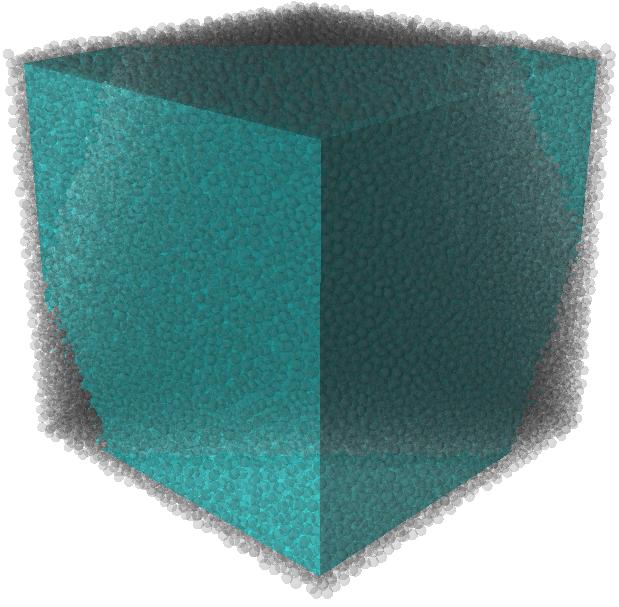}
\caption{Imaginary box positioned inside a random realization}\label{fig8}
\end{figure}
\begin{figure}[!ht]
\centering
\includegraphics[scale=0.15,angle=0]{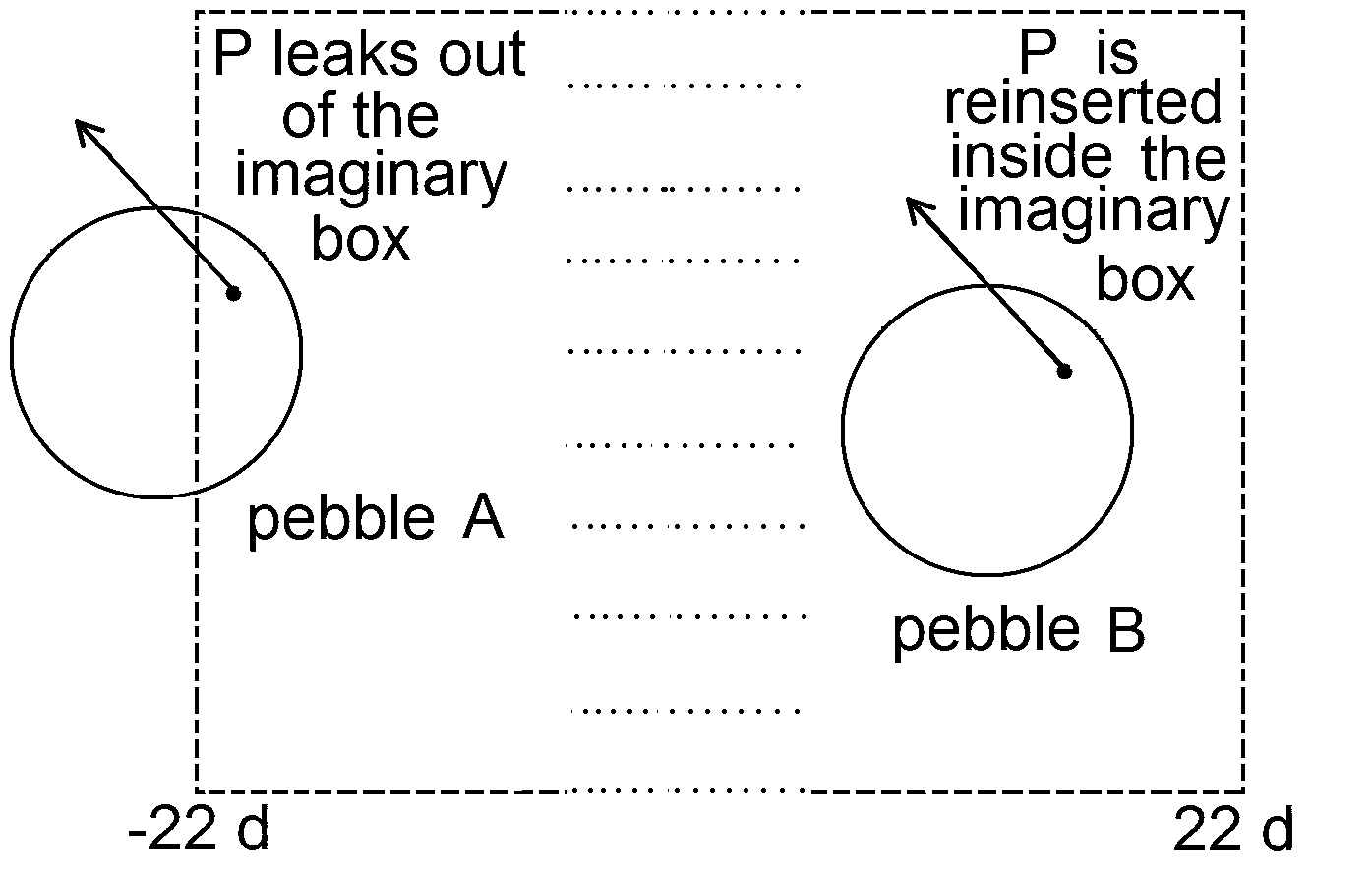}
\caption{Reinsertion of a particle in the imaginary box}\label{fig9}
\end{figure}
\cite{bedenig_68}, we need to consider pebbles that are three to five diameters off-walls in order to have a packing structure that is not influenced by the walls and by the bottom.
For each random realization, we found that the fluctuations in packing fractions ceased being significant around a distance of two diameters from the boundaries, as was the case in the experiment performed in 
\cite{lieberoth_80}. Nevertheless, according to the work in 
\cite{bedenig_68}, one needs to consider pebbles that are three to five diameters off-walls in order to have a packing structure that is not influenced by the walls and by the bottom. Thus, in the Monte Carlo simulations performed, we allow neutrons to travel only inside the imaginary box with side $L^*=44$d depicted in Figure \ref{fig8}.

The difference in the algorithm is in dealing with neutrons that leak out of this imaginary box. Let us assume that the center of the box is at the origin, and let us consider a particle $P$ that 
had its last collision at point $(x_0,y_0,z_0)$ inside a pebble A  and that leaks out of the imaginary box through the plane $x=-22$d. 
 First, defining the coordinates of the center of the pebble A as $(x_a,y_a,z_a)$, we locate the pebble B with center at $(x_b,y_b,z_b)$ closest to the point $(-x_a-\text{d},y_a,z_a)$.
 Then, we \textit{reinsert} $P$ into the system at the point $(x_b+x_0-x_a,y_b+y_0-y_a,z_b+z_0-z_a)$, as shown in Figure \ref{fig9}.
 Finally, we shift the whole system so that now the coordinates of the center of the box are $(x_0,0,0)$, and proceed with the history of the particle. A similar process is used if the particle leaks through any of the other walls; we repeat this reinsertion and shifting procedure as many times as necessary. In other
words, particles are traveling in an infinite ``quasi-periodic" structure; here,
\begin{table}[!ht]
\small
\centering
\caption{Monte Carlo results in random structures}\label{tab3}
\begin{tabular}{||c||c|c|c|c||c|c|c|c||}\hline\hline
&\multicolumn{4}{|c||}{\bf{Problem 1}}&\multicolumn{4}{|c||}{\bf{Problem 2}}\\
\cline{2-9}
&&&&&&&&\\
 & $\bl s\bg/$d & $\bl s^2\bg/$d & $\bl x^2\bg/$d & $\bl z^2\bg/$d &$\bl s\bg/$d & $\bl s^2\bg/$d & $\bl x^2\bg/$d & $\bl z^2\bg/$d \\
&&&&&&&&\\
\hline\hline
{\footnotesize Ensemble} &&&&&&&&\\
{\footnotesize Average}&
 1.7053 & 6.2898 & 209.56 & 210.01 &
 0.8527 & 1.7016 & 224.18 & 224.78\\
\hline
{\footnotesize Statistical} &&&&&&&&\\
{\footnotesize Error} &
0.095\% & 0.211\% & 0.201\% & 0.200\% &
0.080\% & 0.195\% & 0.172\% & 0.179\% 
  \\
\hline
\hline
\end{tabular}
\end{table}
\begin{figure}[!ht]
\centering
\includegraphics[scale=0.5,angle=0]{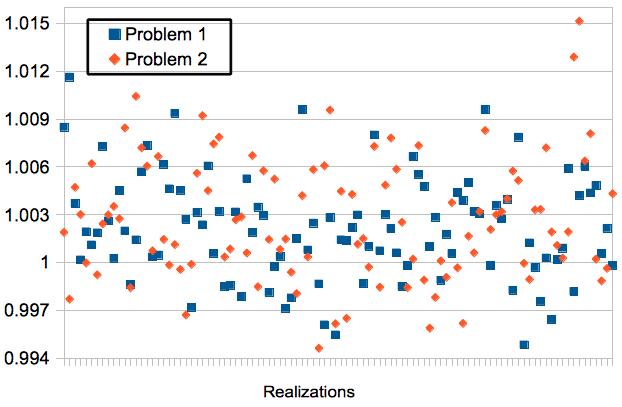} 
\caption{Ratios $\bl z^2\bg/\bl x^2\bg$ in 100 random realizations}\label{fig10}
\end{figure}
we use the term ``quasi-periodic" because the system is not always shifted by the same values in a given direction.

We have simulated the histories of $10^6$ neutrons in each realization of the random system. The statistical error in each realization was found to be 
(with 97.5$\%$ confidence) less than 0.056$\%$ in Problem 1 and 0.032$\%$ in Problem 2 for all values of $\bl s\bg$, $\bl s^2\bg$, $\bl x^2\bg=\bl y^2\bg$, and $\bl z^2\bg$. The ensemble-averaged Monte Carlo results and the statistical errors (with 95$\%$ confidence) are given in Table \ref{tab3}.
Each single realization presents a small anisotropic effect (see Figure \ref{fig10}). This anisotropy is not consistent: (i) we see that $\bl x^2\bg > \bl z^2\bg$ in about 25\% of the realizations;
and (ii) the ensemble-averaged values of $\bl x^2\bg$ and $\bl z^2\bg$ are less than 0.27\% apart in both problems. These facts indicate that, on average, neutrons tend to travel further in the vertical direction than in the horizontal direction; however, this extra distance is {\em very} small. 

Furthermore, we remark that we are working with an average packing fraction of 59.34\%, which is about 2\% smaller than the generally assumed average packing fraction in a PBR core (60\%-62\%). Thus, one expects the small anisotropic effects found in each realization to be even smaller in the interior of real PBR cores, since the pebbles are packed more closely and the void fraction is reduced.

It is not unreasonable to assume that, if enough realizations of the system are simulated, one should find that $\bl z^2\bg/\bl x^2\bg$ approaches a value of $\approx$1. This indicates that, at least for fuel pebbles in the interior of the system, it should not be necessary to worry about anisotropic diffusion. However, the diffusion of neutrons that are born in pebbles positioned close to the walls \textit{is} anisotropic \cite{vasques_13}; we do not address this issue here.
\begin{table}[!ht]
\small
\centering
\caption{Expressions to compute the diffusion coefficients for each model}\label{tab4}
\begin{tabular}{||c||c|c|c|c|c|c||}\hline\hline
& Monte & Atomic & Behrens & Lieberoth & Isotropic & New\\
& Carlo & Mix & Correction & Correction &  GLBE & GLBE\\
\hline\hline
D$_x$ & Eq.\ \eqref{2.22} &Eq.\ \eqref{2.7b} &Eq.\ \eqref{2.8} &Eq.\ \eqref{2.11} &Eq.\ \eqref{2.14b} &Eq.\ \eqref{2.13b}\\
\hline
D$_z$ &Eq.\ \eqref{2.22} &Eq.\ \eqref{2.7b} &Eq.\ \eqref{2.8} &Eq.\ \eqref{2.11} &Eq.\ \eqref{2.14b} &Eq.\ \eqref{2.13c}\\
\hline
\hline
\end{tabular}
\end{table}
\begin{table}[!ht]
\small
\centering
\caption{Diffusion coefficients in crystal structures: Problem 1.}\label{tab5}
\begin{tabular}{||c||c|c||c|c|c||c||c|c||}\hline\hline
& \multicolumn{2}{c||}{Monte} & \multicolumn{3}{c||}{Atomic Mix} & ``Old" & \multicolumn{2}{c||}{New}\\
& \multicolumn{2}{c||}{Carlo} & \multicolumn{3}{c||}{and Corrections} & GLBE & \multicolumn{2}{c||}{GLBE}\\
\cline{2-9}
&&&&&&&&\\
$\varepsilon/\underline d$ &D$_x^{mc}$&D$_z^{mc}$&D$^{am}$&D$^{B}$&D$^{L}$ & D$^{iso}$  &D$_{x}^{gt}$ & D$_z^{gt}$\\
&&&&&&&&\\
\hline\hline
0.000
&0.4635
&0.4628
&0.4502
&0.4705
&0.4626
&0.4646
&0.4646
&0.4646\\
\hline
0.025
&0.4830
&0.4841
&0.4676
&0.4903
&0.4833
&0.4842
&0.4836
&0.4855\\
\hline
0.050
&0.5016
&0.5037
&0.4841
&0.5093
&0.5031
&0.5039
&0.5027
&0.5063
\\ \hline
0.075
&0.5210
&0.5250
&0.5004
&0.5282
&0.5229
&0.5236
&0.5219
&0.5271
\\ \hline
0.100
&0.5404
&0.5456
&0.5162
&0.5467
&0.5423
&0.5432
&0.5410
&0.5476
\\ \hline
0.125
&0.5579
&0.5660
&0.5313
&0.5646
&0.5610
&0.5624
&0.5598
&0.5677
\\ \hline
0.150
&0.5765
&0.5873
&0.5463
&0.5825
&0.5797
&0.5814
&0.5784
&0.5875
\\ \hline
0.175
&0.5936
&0.6036
&0.5603
&0.5993
&0.5973
&0.5998
&0.5966
&0.6063
\\ \hline
0.200
&0.6109
&0.6238
&0.5743
&0.6160
&0.6149
&0.6176
&0.6142
&0.6244
\\ \hline
0.225
&0.6282
&0.6388
&0.5861
&0.6304
&0.6299
&0.6345
&0.6310
&0.6416
\\ \hline
0.250
&0.6443
&0.6534
&0.5977
&0.6444
&0.6446
&0.6504
&0.6469
&0.6573
\\ \hline
0.275
&0.6586
&0.6686
&0.6077
&0.6567
&0.6574
&0.6651
&0.6618
&0.6719
\\ \hline
0.300
&0.6723
&0.6814
&0.6170
&0.6680
&0.6693
&0.6783
&0.6752
&0.6845
\\ \hline
0.325
&0.6834
&0.6919
&0.6253
&0.6782
&0.6800
&0.6897
&0.6871
&0.6950
\\ \hline
0.350
&0.6926
&0.6982
&0.6312
&0.6854
&0.6875
&0.6991
&0.6970
&0.7034
\\ \hline
0.375
&0.7012
&0.7051
&0.6355
&0.6907
&0.6931
&0.7059
&0.7046
&0.7086
\\ \hline
0.400
&0.7050
&0.7058
&0.6378
&0.6935
&0.6960
&0.7095
&0.7089
&0.7105
\\ \hline
0.425
&0.7069
&0.7052
&0.6374
&0.6930
&0.6955
&0.7098
&0.7102
&0.7091
\\ \hline
0.450
&0.7054
&0.6984
&0.6346
&0.6896
&0.6919
&0.7059
&0.7072
&0.7031
\\ \hline
0.475
&0.6968
&0.6879
&0.6305
&0.6845
&0.6867
&0.6976
&0.6999
&0.6931
\\ \hline
0.500
&0.6847
&0.6725
&0.6212
&0.6732
&0.6747
&0.6846
&0.6877
&0.6785
\\ \hline
0.525
&0.6672
&0.6565
&0.6089
&0.6581
&0.6590
&0.6661
&0.6697
&0.6589
\\ \hline
0.550
&0.6432
&0.6313
&0.5931
&0.6389
&0.6388
&0.6419
&0.6457
&0.6344
\\ \hline
0.575
&0.6139
&0.6024
&0.5711
&0.6122
&0.6108
&0.6114
&0.6149
&0.6045
\\ \hline
0.600
&0.5749
&0.5664
&0.5423
&0.5776
&0.5747
&0.5738
&0.5763
&0.5688
\\ \hline
0.625
&0.5274
&0.5229
&0.5048
&0.5334
&0.5283
&0.5278
&0.5285
&0.5266\\
\hline
\hline
\end{tabular}
\end{table}

\vspace{10pt}
\noindent
{\bf 5. ESTIMATED DIFFUSION COEFFICIENTS}
\setcounter{section}{5}
\setcounter{equation}{0} 
\vspace{10pt}

In this section, we compare the diffusion coefficients obtained numerically with those estimated by the different models presented in Section 2.
\begin{table}[!ht]
\small
\centering
\caption{Diffusion coefficients in crystal structures: Problem 2.}\label{tab6}
\begin{tabular}{||c||c|c||c|c|c||c||c|c||}\hline\hline
& \multicolumn{2}{c||}{Monte} & \multicolumn{3}{c||}{Atomic Mix} & ``Old" & \multicolumn{2}{c||}{New}\\
& \multicolumn{2}{c||}{Carlo} & \multicolumn{3}{c||}{and Corrections} & GLBE & \multicolumn{2}{c||}{GLBE}\\
\cline{2-9}
&&&&&&&&\\
$\varepsilon/\underline d$ &D$_x^{mc}$&D$_z^{mc}$&D$^{am}$&D$^{B}$&D$^{L}$ & D$^{iso}$  &D$_{x}^{gt}$ & D$_z^{gt}$\\
&&&&&&&&\\
\hline\hline
0.000
&0.2386
&0.2385
&0.2251
&0.2455
&0.2385
&0.2402
&0.2402
&0.2402
\\ \hline
0.025
&0.2491
&0.2513
&0.2338
&0.2565
&0.2508
&0.2516
&0.2510
&0.2528
\\ \hline
0.050
&0.2598
&0.2650
&0.2420
&0.2672
&0.2627
&0.2633
&0.2621
&0.2656
\\ \hline
0.075
&0.2712
&0.2771
&0.2502
&0.2780
&0.2747
&0.2751
&0.2734
&0.2785
\\ \hline
0.100
&0.2817
&0.2892
&0.2581
&0.2886
&0.2864
&0.2870
&0.2848
&0.2914
\\ \hline 
0.125
&0.2931
&0.3025
&0.2656
&0.2989
&0.2979
&0.2989
&0.2963
&0.3041
\\ \hline
0.150
&0.3037
&0.3139
&0.2732
&0.3093
&0.3094
&0.3107
&0.3077
&0.3165
\\ \hline
0.175
&0.3148
&0.3256
&0.2802
&0.3191
&0.3203
&0.3222
&0.3191
&0.3284
\\ \hline
0.200
&0.3260
&0.3366
&0.2871
&0.3289
&0.3312
&0.3335
&0.3302
&0.3401
\\ \hline
0.225
&0.3360
&0.3476
&0.2931
&0.3373
&0.3406
&0.3443
&0.3410
&0.3509
\\ \hline
0.250
&0.3459
&0.3568
&0.2989
&0.3456
&0.3498
&0.3546
&0.3514
&0.3611
\\ \hline
0.275
&0.3558
&0.3667
&0.3039
&0.3528
&0.3578
&0.3642
&0.3611
&0.3703
\\ \hline
0.300
&0.3635
&0.3745
&0.3085
&0.3595
&0.3653
&0.3729
&0.3700
&0.3785
\\ \hline
0.325
&0.3722
&0.3811
&0.3127
&0.3655
&0.3720
&0.3805
&0.3781
&0.3853
\\ \hline
0.350
&0.3788
&0.3842
&0.3156
&0.3698
&0.3768
&0.3868
&0.3849
&0.3905
\\ \hline
0.375
&0.3841
&0.3883
&0.3177
&0.3729
&0.3803
&0.3913
&0.3901
&0.3938
\\ \hline
0.400
&0.3880
&0.3892
&0.3189
&0.3746
&0.3821
&0.3939
&0.3934
&0.3949
\\ \hline
0.425
&0.3890
&0.3875
&0.3187
&0.3743
&0.3818
&0.3940
&0.3943
&0.3933
\\ \hline
0.450
&0.3870
&0.3813
&0.3173
&0.3723
&0.3796
&0.3915
&0.3927
&0.3889
\\ \hline
0.475
&0.3822
&0.3735
&0.3153
&0.3693
&0.3762
&0.3858
&0.3879
&0.3816
\\ \hline
0.500
&0.3749
&0.3646
&0.3106
&0.3625
&0.3687
&0.3769
&0.3798
&0.3713
\\ \hline
0.525
&0.3635
&0.3516
&0.3045
&0.3537
&0.3588
&0.3646
&0.3680
&0.3579
\\ \hline
0.550
&0.3484
&0.3356
&0.2966
&0.3423
&0.3462
&0.3488
&0.3523
&0.3418
\\ \hline
0.575
&0.3293
&0.3174
&0.2855
&0.3266
&0.3287
&0.3292
&0.3324
&0.3227
\\ \hline
0.600
&0.3052
&0.2968
&0.2711
&0.3065
&0.3063
&0.3054
&0.3078
&0.3008
\\ \hline
0.625
&0.2758
&0.2736
&0.2524
&0.2810
&0.2779
&0.2772
&0.2778
&0.2759\\
\hline
\hline
\end{tabular}
\end{table}
Table \ref{tab4} summarizes the expressions to compute
the different diffusion coefficients.
The Monte Carlo diffusion coefficients D$^{mc}_x$ and D$^{mc}_z$ are computed using the numerically calculated mean free path $\bl s\bg$ and mean-squared displacements $\bl x^2\bg$ and $\bl z^2\bg$. The values obtained with the generalized theory (D$^{iso}$, D$^{gt}_x$, and D$^{gt}_z$) use the numerical results for $s^2_{\bm\Omega}$, $\bl s\bg$, and $\bl s^2\bg$. The atomic mix estimates (D$^{am}$, D$^B$, and D$^L$) use the Monte Carlo ensembe-averaged packing fraction $\bl \Gamma^* \bg = 0.5934$ for the random structures and the packing fractions depicted in Figure \ref{fig4} for the crystal structures.   

To investigate the accuracy of the theoretical estimates, we define the relative differences (errors) between the theoretical and Monte Carlo diffusion coefficients in the direction $u$ as
\begin{equation}\label{5.1}
\text{error}_{u} = \frac{|\text{D}^{model}_u-\text{D}^{mc}_u|}{\text{D}^{mc}_u},
\end{equation}
where $u$ can be replaced by $x=y$ and $z$; {\em mc} stands for the results obtained by monte carlo; and {\em model} represents the model being compared.

\vspace{5pt}
\noindent
{\bf 5.1 Estimates for Crystal Structures}
\vspace{5pt}

The estimates for the diffusion coefficients in crystal structures are given in Tables \ref{tab5} (Problem 1) and \ref{tab6} (Problem 2), for
different values of $\varepsilon$ considered.
\begin{figure}[!ht]
\centering
\includegraphics[scale=0.43,angle=0]{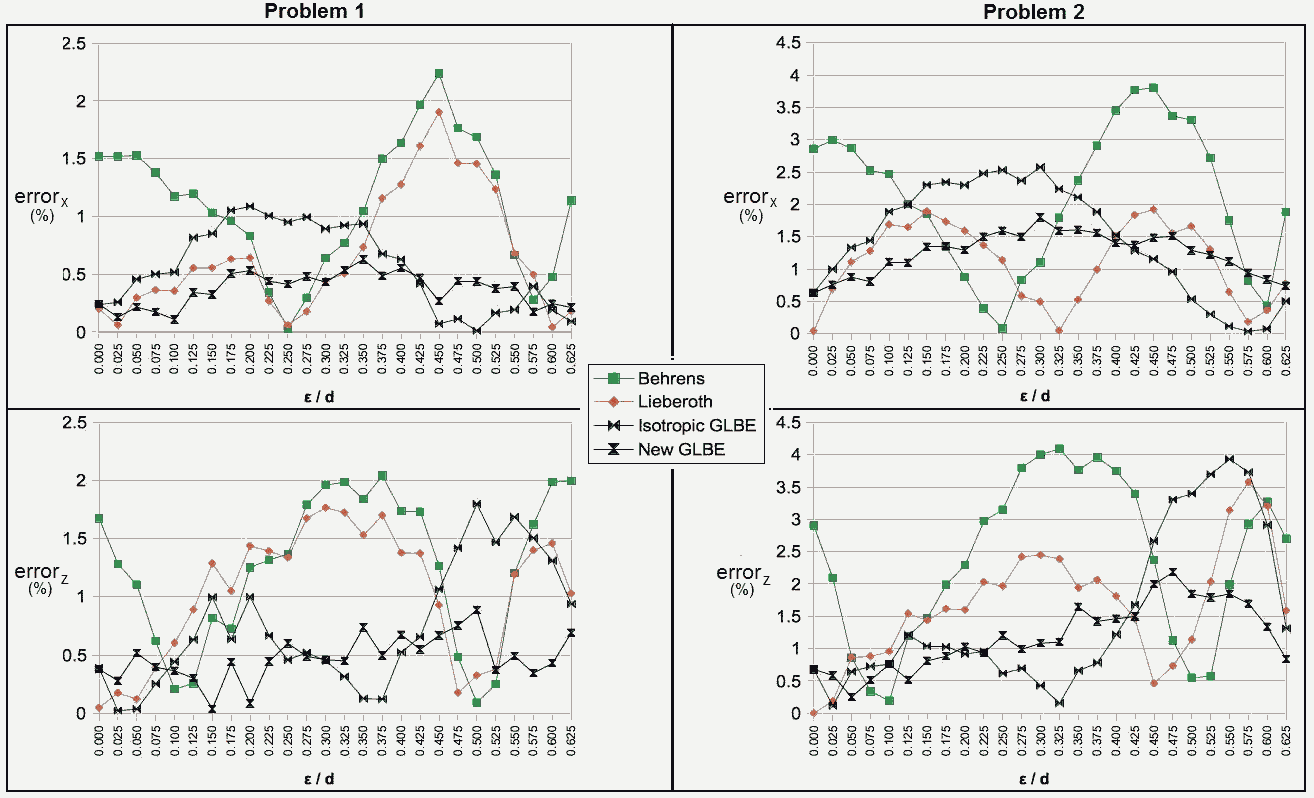}
\caption{Percent relative errors in crystal structures for different values of $\varepsilon$}\label{fig11}
\end{figure}  
The results obtained with classic atomic mix consistently underestimate the diffusion coefficients in both directions, differing from the numerical results by large amounts: up to 10\% in Problem 1 and 18\% in Problem 2. On the other hand, the estimates computed with the other models yield much smaller differences.

The relative errors obtained by each model [as defined in Eq.\ \eqref{5.1}] are plotted in Figure \ref{fig11}. The Behrens correction presents the largest errors, despite being relatively accurate. The correction suggested by Lieberoth and the results obtained with the Isotropic GLBE perform similarly. Since D$^{iso}>$D$^L$ for almost every choice of $\varepsilon$, the accuracy of both methods alternates accordingly to the anisotropy encountered in each case. Once again, we point out that these methods give {\em isotropic} diffusion coefficients, which do not model the anisotropic behavior of neutrons found in these systems.

However, the new GLBE {\em correctly identifies} the anisotropic
 behavior in every case; that is, it yields D$_x^{gt}>$D$_z^{gt}$ when D$_x^{mc}>$D$_z^{mc}$, and D$_x^{gt}<$D$_z^{gt}$ when D$_x^{mc}<$D$_z^{mc}$. It is an improvement over the other methods, with a maximum error of 0.89\% in Problem 1 (against 2.24\% for D$^B$, 1.90\% for D$^L$, and 1.80\% for D$^{iso}$), and 2.18\% in Problem 2 (against 4.09\% for D$^B$, 3.57\% for D$^L$, and 3.93\% for D$^{iso}$).
 
\vspace{5pt}
\noindent
{\bf 5.2 Estimates for Random Structures}
\vspace{5pt}

The estimates for the diffusion coefficients in random structures are given in Table \ref{tab7}, as well as the relative errors of the theoretical estimates compared to
\begin{table}[!ht]
\small
\centering
\caption{Diffusion coefficients in random structures}\label{tab7}
\begin{tabular}{||cc||c|c||c|c|c||c||c|c||}\hline\hline
&& \multicolumn{2}{c||}{Monte} & \multicolumn{3}{c||}{Atomic Mix} & ``Old" & \multicolumn{2}{c||}{New}\\
&& \multicolumn{2}{c||}{Carlo} & \multicolumn{3}{c||}{and Corrections} & GLBE & \multicolumn{2}{c||}{GLBE}\\
\cline{3-10}
&&&&&&&&&\\
\multicolumn{2}{||c||}{{\bf Problem}} &D$_x^{mc}$&D$_z^{mc}$&D$^{am}$&D$^{B}$&D$^{L}$ & D$^{iso}$  &D$_{x}^{gt}$ & D$_z^{gt}$\\
&&&&&&&&&\\
\hline\hline
\multicolumn{1}{||c|}{$\mathbf{\cdot}$}&{\footnotesize Diffusion} &&&&&&&&\\
\multicolumn{1}{||c|}{$\mathbf{\cdot}$} &{\footnotesize Coeff.}&
0.6144
&0.6157
&0.5617
&0.6009
&0.5990
&0.6147
&0.6146
&0.6150\\
\cline{2-10}
\multicolumn{1}{||c|}{{\bf 1}} &{\footnotesize error$_x$} &&&&&&&&\\
\multicolumn{1}{||c|}{$\mathbf{\cdot}$} &{\footnotesize (\%)} &
-
& -
&8.580
&2.201
&2.506
&0.049
&0.029
& -\\
\cline{2-10}
\multicolumn{1}{||c|}{$\mathbf{\cdot}$} &{\footnotesize error$_z$} &&&&&&&&\\
\multicolumn{1}{||c|}{$\mathbf{\cdot}$} &{\footnotesize (\%)} &
-
& -
&8.776
&2.411
&2.716
&0.166
& -
&0.126
\\
\hline
\hline
\multicolumn{1}{||c|}{$\mathbf{\cdot}$}&{\footnotesize Diffusion} &&&&&&&&\\
\multicolumn{1}{||c|}{$\mathbf{\cdot}$} &{\footnotesize Coeff.}
&0.3286
&0.3295
&0.2809
&0.3200
&0.3214
&0.3326
&0.3324
&0.3329\\
\cline{2-10}
\multicolumn{1}{||c|}{{\bf 2}} &{\footnotesize error$_x$} &&&&&&&&\\
\multicolumn{1}{||c|}{$\mathbf{\cdot}$} &{\footnotesize (\%)} &
-
& -
&14.542
&2.617
&2.214
&1.204
&1.154
& -\\
\cline{2-10}
\multicolumn{1}{||c|}{$\mathbf{\cdot}$} &{\footnotesize error$_z$} &&&&&&&&\\
\multicolumn{1}{||c|}{$\mathbf{\cdot}$} &{\footnotesize (\%)} &
-
& -
&14.771
&2.877
&2.475
&0.934
& -
&1.034
\\
\hline
\hline
\end{tabular}
\end{table}
the ensemble-averaged Monte Carlo results.
Similarly to the results in crystal structures, classic atomic mix performs poorly. The estimates computed by the Behrens and Lieberoth corrections underestimate the diffusion coefficients, with relative errors ranging from 2.2\% to 2.9\%. More importantly, the estimates computed with the GLBE are a clear improvement over the other methods, presenting an excellent level of accuracy. They outperform atomic mix and its corrections by one order of magnitude in Problem 1 and by a factor of 2 in Problem 2.  

As with the crystal structures, the new GLBE {\em correctly identifies} the anisotropic behavior in both random problems, yielding a larger diffusion coefficient in the vertical direction. It is slightly more accurate than the Isotropic GLBE in Problem 1, and as accurate as the Isotropic GLBE in Problem 2. 

\vspace{10pt}
\noindent
{\bf 6. DISCUSSION}
\setcounter{section}{5}
\setcounter{equation}{0} 
\vspace{10pt}

In this work, we have investigated anisotropic diffusion of neutrons in the interior of model pebble bed reactor (PBR) cores, in which pebbles are arranged in crystal or random structures. Using Monte Carlo simulations, we have found clear anisotropic behavior in the crystal structures; neutrons travel longer in certain directions, according to the geometry and the packing fraction of the system. We have also found a {\em very} small anisotropic behavior in the ensemble-averaged random structures. 

We have used the diffusion approximation of a new generalized linear Boltzmann equation (new GLBE) to estimate anisotropic diffusion coefficients that can capture this anisotropic behavior. This new theory utilizes a non-classical form of the Boltzmann equation in which the locations of the scattered centers (pebbles) are correlated and the distance-to-collision is not given by an exponential. In order to successfully apply this theory, we need to estimate the mean and the (angular-dependent) mean-squared free paths of neutrons; we do this numerically. This extra information is all microscopic in nature; it is not a closure relation. We have shown that, at least for problems of the pebble bed kind, this new approach can correctly identify even very small anisotropic diffusion, which is a feature not encountered in any of the standard approaches currently in use.

We compare the accuracy of the new GLBE estimates for the diffusion coefficients against other methods: the atomic mix model and two of its diffusion corrections, and the standard GLBE. These methods yield isotropic diffusion coefficients; that is, they cannot capture any anisotropic behavior present in the systems. Nevertheless, the results obtained with the standard GLBE and with both corrections to atomic mix present a good level of accuracy.

We find that the new GLBE is generally an improvement over the other methods in estimating the diffusion coefficients in crystal structures. It correctly predicts the anisotropic behavior of every system, and its maximum relative errors are about half as big as the ones obtained with the other methods.

More importantly, for the random structures, the new and standard GLBE models were shown to be a great improvement over the other methods, with relative errors one order of magnitude smaller in one problem and about 50\% smaller in the other. The new GLBE slightly outperforms the standard GLBE in one problem, and has similar accuracy in the other. However, once again, the new GLBE is capable of detecting even the very small anisotropic diffusion found in these random systems, which the standard GLBE cannot do.

The GLBE is more costly to simulate than the atomic mix approximation. Atomic mix only requires the cross sections of the constituent materials and their volume fractions to be known. The GLBE requires much more detailed information, which must be obtained by constructing realizations of the random system and developing an accurate estimate of the ensemble-averaged distribution function for distance-to-collision.
However, the GLBE preserves important statistical properties of the original random system, such as the ensemble-averaged probability distribution function of the distance-to-collision. The new GLBE preserves the more general, angular-dependent version of these quantities, which makes it a systematically more accurate alternative to the current methods.

We have also shown that, in the case of diffusion in the interior of random PBR systems, one should not have to worry about anisotropic diffusion. The anisotropic effect was found to be very small for the tested packing fractions, and it is likely to be even smaller for higher packing fractions encountered in actual cores. Nevertheless, diffusion {\em is} anisotropic for neutrons that are born in pebbles close to the boundaries of the core (neutrons tend to travel longer distances in directions parallel to the boundary), and the new GLBE theory can be used to capture this behavior.

In future work, we intend to extend these results for problems with anisotropic scattering and energy-dependence. Although it is computationally expensive, an algorithm to obtain $\Sigma_t(\bm\Omega,s)$ is straightforward; we intend to use it to generate results that can be applied to the new GLBE in order to estimate the criticality eigenvalue $k$. Monte Carlo benchmark results for $k$ in the heterogeneous core will need to be developed; we also intend to do this. 


\vspace{14pt}
\setlength{\baselineskip}{14pt}
\renewcommand{\section}[2]{{\bf \vspace{10pt} \noindent REFERENCES\\}}

\end{document}